\documentclass[nonacm,sigplan]{acmart}

\usepackage{hyperref}
\usepackage{hyperxmp}

\usepackage{tikz}
\usepackage{xcolor}
\usepackage[normalem]{ulem}
\usepackage{amsmath}
\usepackage{multirow}
\usepackage{diagbox}
\usepackage{makecell}
\usepackage{color, soul}
\usepackage{titlesec}
\usepackage[capitalise]{cleveref}
\usepackage{multicol} 
\usepackage{float}
\usepackage[linesnumbered,ruled,vlined]{algorithm2e}

\newcommand{\frameworkname}{PowerMove}

\begin{document}

\title{\frameworkname: Optimizing Compilation for Neutral Atom Quantum Computers with Zoned Architecture}

\author{Jixuan Ruan\textsuperscript{*}}
\email{j3ruan@ucsd.edu}
\affiliation{
    \institution{University of California}
    \city{San Diego}
    \country{USA}
}
\author{Xiang Fang\textsuperscript{*}}
\email{x8fang@ucsd.edu}
\affiliation{
    \institution{University of California}
    \city{San Diego}
    \country{USA}
}
\author{Hezi Zhang}
\email{hez019@ucsd.edu}
\affiliation{
    \institution{University of California}
    \city{San Diego}
    \country{USA}
}
\author{Ang Li}
\email{ang.li@pnnl.gov}
\affiliation{
    \institution{Pacific Northwest National Laboratory}
    \city{Richland}
    \country{USA}
}
\author{Travis Humble}
\email{humblets@ornl.gov}
\affiliation{
    \institution{Oak Ridge National Laboratory}
    \city{Oak Ridge}
    \country{USA}
}
\author{Yufei Ding}
\email{yufeiding@ucsd.edu}
\affiliation{
    \institution{University of California}
    \city{San Diego}
    \country{USA}
}
\begin{abstract}
    Neutral atom-based quantum computers (NAQCs) have recently emerged as promising candidates for scalable quantum computing, largely due to their advanced hardware capabilities, particularly qubit movement and the zoned architecture (ZA). However, fully leveraging these features poses significant compiler challenges, as it requires addressing complexities across gate scheduling, qubit allocation, qubit movement, and inter-zone communication. In this paper, we present~\emph{\frameworkname}, an efficient compiler for NAQCs that enhances the qubit movement framework while fully integrating the advantages of ZA. By recognizing and leveraging the interdependencies between these key aspects, \frameworkname~unlocks new optimization opportunities, significantly enhancing both scalability and fidelity. Our evaluation demonstrates an improvement in fidelity by several orders of magnitude compared to the state-of-the-art methods, with execution time improved by up to 3.46x and compilation time reduced by up to 213.5x. We will open-source our code later to foster further research and collaboration within the community.
\end{abstract}

\maketitle 
\pagestyle{plain} 

\section{Introduction}\label{sec:intro}

\begin{figure*}[!ht]
    \centering
    \includegraphics[width=1\textwidth]{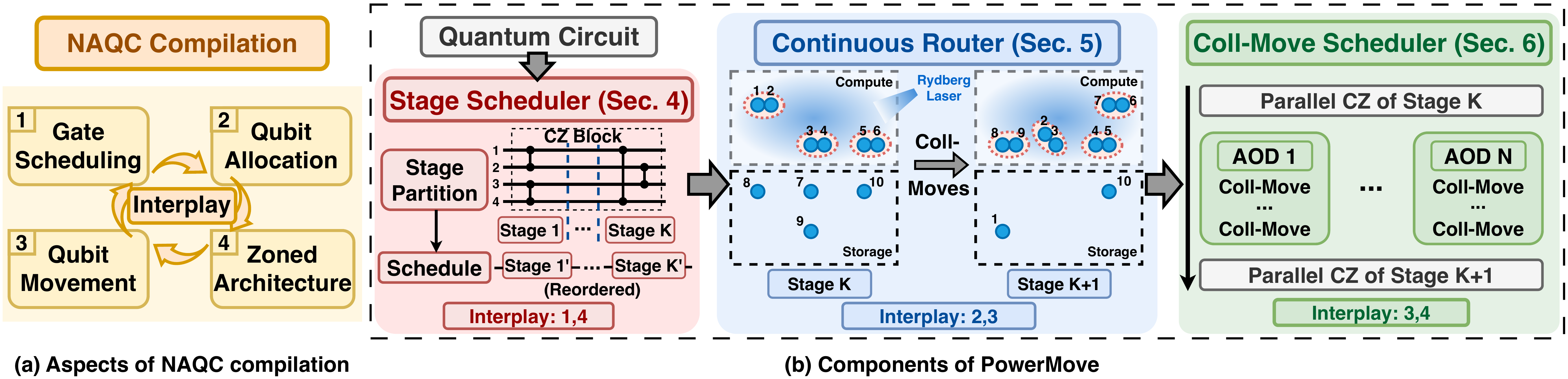}
    \caption{(a) Four key aspects of the NAQC compilation problem. (b) Overview of the \frameworkname~framework. The design of each component is based on the interplay of multiple aspects of the problem.}
    \Description{Intro.}
    \label{fig:introfig}
\end{figure*}

Quantum computing (QC) is swiftly evolving from a theoretical concept into a tangible reality, with significant advancements across various platforms over the past few decades~\cite{arute2019quantum, huang2020superconducting, bravyi2022future, bruzewicz2019trapped, chen2023benchmarking, kok2007linear, bourassa2021blueprint, bluvstein2024logical, wurtz2023aquila}. Given the rapid progress and the unique strengths of each platform, competition among these platforms is expected to persist for the foreseeable future.

In recent years, \emph{neutral atom-based quantum computer (NAQC)}~\cite{ebadi2021quantum, graham2022multi, wurtz2023aquila, bluvstein2024logical} have emerged as a strong candidate in the QC landscape, due to its unique hardware advantages. These include impressive scalability (supporting up to 6100 qubits) ~\cite{manetsch2024tweezer, ebadi2022quantum, wurtz2023aquila, ebadi2021quantum}, long coherence times of several seconds~\cite{evered2023high, young2020half, barnes2022assembly}, and high-fidelity operations~\cite{bluvstein2024logical, levine2022dispersive, bluvstein2022quantum, levine2019parallel, fu2022high, evered2023high}, with single-qubit rotations and two-qubit CZ gates achieving fidelities of up to 99.99\% and 99.5\%, respectively.

Beyond these fundamental features, NAQC offers the compelling ability to move qubits collectively using \emph{AOD} under some constraints~\cite{bluvstein2022quantum}, enabling non-local connectivity and dynamic layouts that facilitate parallel execution of CZ gates. Once interacting qubit pairs are brought close together, a global Rydberg laser~\cite{levine2019parallel, evered2023high} can perform CZ gates between each pair. This dynamic control has further led to the development of the Zoned Architecture (ZA)~\cite{bluvstein2024logical}, which divides the system into distinct zones for specific tasks, such as storage and computation (e.g., CZ gates), with qubits shuttled between zones as needed. Similar to classical architectures with separate memory and processing units, this design may enhance overall system performance. For example, non-interacting qubits can be moved to a storage zone, where they are preserved with negligible decoherence and are protected from excitation errors induced by the Rydberg laser.



Several compilers have been developed~\cite{wang2024atomique, wang2023q, tan2022qubit, tan2024compiling, bochen2024compilation} to harness NAQC’s qubit movement capabilities. While these approaches have made remarkable progress, they still fall short of effectively exploiting the flexibility that qubit movement offers. They either settle for a partially fixed layout~\cite{wang2024atomique, wang2023q, bochen2024compilation} or struggle to handle fully dynamic layout transitions in a scalable manner~\cite{tan2022qubit, tan2024compiling}. Additionally, the potential of the ZA remains unexplored due to its recent introduction, and the limited use of movement capabilities restricts these approaches from extending to this new setting. Our goal is to unlock the full potential of dynamic layout transitions while integrating ZA to further enhance compilation performance.


To achieve this goal, we first identify four key aspects of the NAQC compilation problem. (1) \emph{Gate scheduling.} CZ gates are grouped into distinct \emph{stages}, where gates within the same stage act on disjoint qubits and can be executed in parallel. (2) \emph{Qubit allocation.} For each stage, qubits must be strategically placed with appropriate spacing to enable desired CZ gates while avoiding unwanted interactions during Rydberg laser excitation. (3) \emph{Qubit movement.} After each stage, qubits are rearranged for the next stage through collective movements, which must comply with specific rules~\cite{bluvstein2024logical, bluvstein2022quantum}. (4) \emph{Zoned architecture (ZA).} During layout transitions, non-interacting qubits should be moved to the storage zone for protection, while interacting qubits must be brought out of storage for computation. 

Handling all of these aspects simultaneously is highly challenging due to the vast design space they create, and we identified this as the core reason for the limitations of previous approaches. Solver-based methods~\cite{tan2022qubit, tan2024compiling} attempt to tackle this space directly, but face scalability issues. Other approaches~\cite{wang2024atomique, wang2023q, bochen2024compilation} decompose the problem into sub-problems, each addressing a single aspect. While this decomposition leads to more efficient solutions, it overlooks crucial optimization opportunities arising from the interdependencies and synergies between these aspects.

We propose a novel NAQC compiler,~\textbf{\frameworkname}, to effectively tackle this vast design space. By fully recognizing and leveraging the interplays between the above key aspects, we unlock new optimization opportunities that significantly enhance both scalability and fidelity. Our solution consists of three key components:

\noindent (1) \emph{Stage Scheduler.} This component utilizes the interplay between gate scheduling and the ZA. We observed that optimizing the execution order of stages can minimize qubit interchange between the computation and storage zones during layout transitions, thereby reducing inter-zone movement overhead (see Sec.~\ref{sec:stage scheduler}).

\noindent (2) \emph{Continuous Router.} This component integrates qubit allocation with qubit movement. While previous methods solely used movement to change qubit allocation for CZ interactions, we recognize that the current qubit allocation can also guide movement decisions for the next stage. This interdependence allows for the simultaneous determination of qubit allocation and movements, enabling \emph{continuous transitions} between desired layouts without relying on intermediate fixed layouts (see Sec.~\ref{sec:conti router}).


\noindent (3) \emph{Coll-Move Scheduler.} This component leverages the interplay between qubit movement and the ZA. By optimizing the execution order of collective movements (Coll-Moves), it maximizes qubit dwell time in the storage zone, thus minimizing decoherence. It also incorporates the scheduling of multiple AOD arrays to further enhance movement parallelism and reduce latency (see Sec.~\ref{sec:Move scheduler}).

To summarize, our contribution of this paper is as follows:
\begin{itemize}
\item We propose \textbf{\frameworkname}, a novel compiler for NAQC that fully leverages qubit movement capabilities while seamlessly integrating the newly developed ZA.

\item We integrate the storage zone for the first time, effectively eliminating excitation errors while minimizing the associated overhead.

\item We introduce a continuous router that enables direct transitions between qubit layouts, significantly reducing movement overhead.

\item We establish a stage scheduler and a Coll-Move scheduler that fully exploit the storage zone’s advantages to minimize decoherence.

\item Our evaluation demonstrates improvements of several orders of magnitude in fidelity, a 1.71x to 3.46x reduction in execution time, and up to a 213.5x reduction in compilation time compared to the current best NAQC compilation framework.
\end{itemize}

\section{Background}\label{sec:background}


This section provides essential background information on NA hardware capabilities and fidelity analysis.

\subsection{NA Hardware Capabilities}\label{subsec: NA hardware cap}
We first introduce \emph{gate operations} and \emph{qubit movement}, which are directly relevant to the fidelity analysis in Sec.~\ref{subsec:fidelity analysis}. We then discuss the \emph{zoned architecture} of NA hardware, which offers new opportunities for compiler optimization.

\begin{figure}[!ht]
    \centering
    \includegraphics[width=0.47\textwidth]{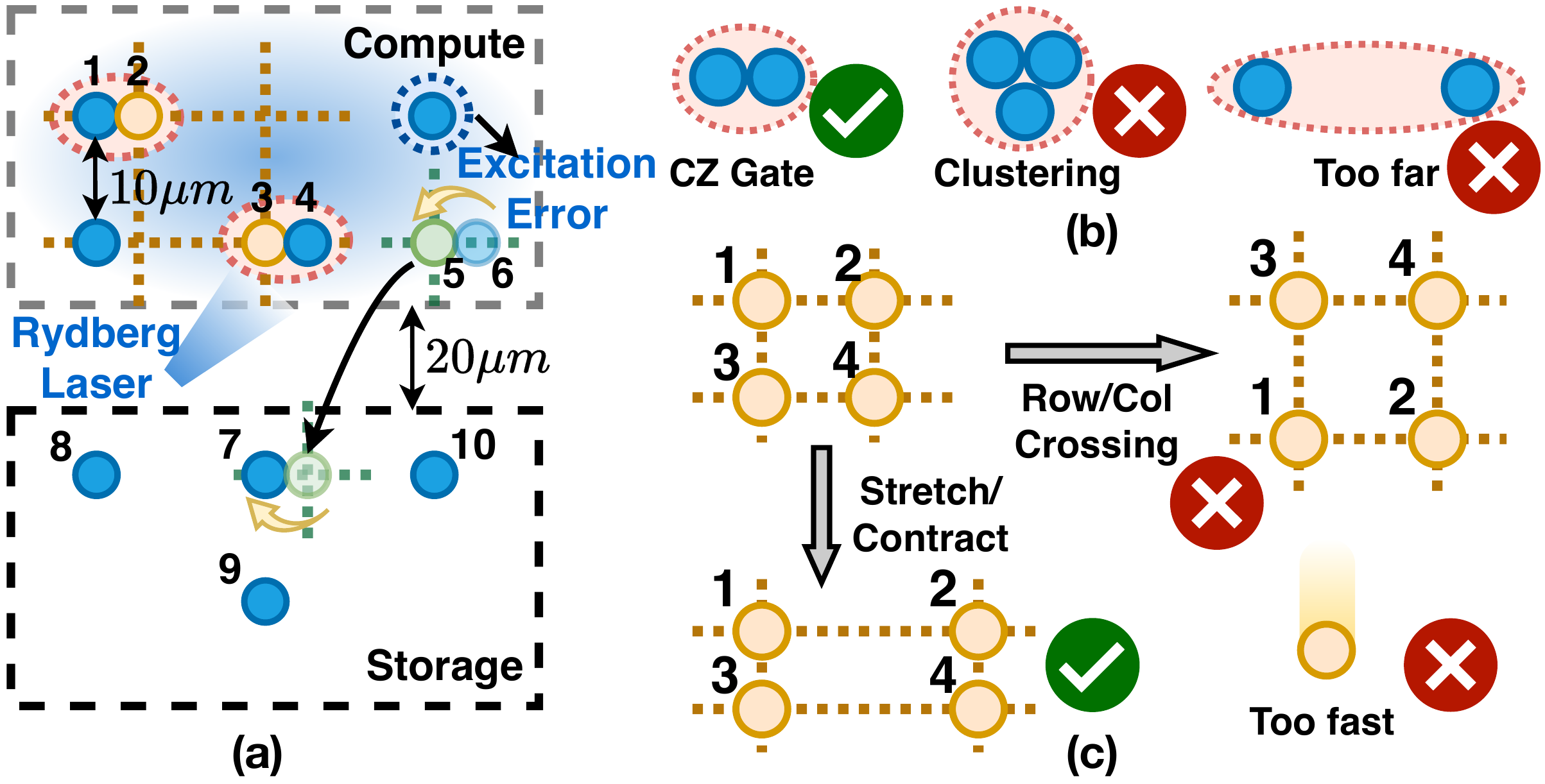}
    \vspace{-6pt}
    \caption{(a) NAQC with zoned architecture. (b) Qubit allocation for CZ gates. (c) Movement constraints of AOD.}
    \Description{Constraints}
    \label{fig:background} 
\end{figure}

\vspace{0.3em}
\noindent\textbf{Gate Operations. }
NAQC supports high-fidelity single-qubit (1Q) rotations and CZ gates, sufficient for universal quantum computing~\cite{boykin1999universal}. \emph{1Q gates} are performed using qubit-specific, parallel Raman pulses, achieving fidelity of $99.99\%$ with duration of $\sim 1 \mu s$~\cite{bluvstein2022quantum, levine2022dispersive, evered2023high, bluvstein2024logical}, allowing for simultaneous execution across the qubit plane. \emph{CZ gates} (specified by the red shaded area in Fig.\ref{fig:background}(a)) are executed by bringing qubits within the Rydberg radius ($r_b \approx 6 \mu m$~\cite{bochen2024compilation}) and applying a global Rydberg excitation~\cite{bluvstein2022quantum, bluvstein2024logical}. Atom pairs within $r_b$ interact via the Rydberg blockade effect~\cite{jaksch2000fast, urban2009observation, levine2019parallel}, while non-interacting qubits must be spaced at least $10\mu m$ apart to avoid clustering that leads to unwanted interactions~\cite{bluvstein2024logical}, as Fig.~\ref{fig:background}(b) shows. Current CZ gate fidelity reaches $99.5\%$ with duration $270ns$~\cite{bluvstein2024logical}, enabling parallel gate execution on distinct qubits provided that they are wisely positioned. However, non-interacting qubits still experience a fidelity reduction to $99.75\%$ during excitation~\cite{bochen2024compilation}, specified by the dotted blue circle in the computation zone in Fig.~\ref{fig:background}(a).


\vspace{0.3em}
\noindent\textbf{Qubit Movement.} Qubit movement is controlled by two types of optical traps~\cite{beugnon2007two}: (1) \emph{static traps} generated by a spatial light modulator (SLM)~\cite{ebadi2021quantum, scholl2021quantum}, and (2) \emph{mobile traps} generated by a crossed 2D acousto-optic deflector (AOD)~\cite{bluvstein2022quantum}, represented by blue and yellow (or green) dots in Fig.~\ref{fig:background}, respectively. These traps are typically arranged in a 2D lattice array. By \emph{transferring} qubits from static to mobile traps and collectively moving them to desired locations, dynamic layout reconfiguration is achieved during computation to enable CZ interactions. The \emph{transfer} process between SLM and AOD traps has a fidelity of $99.9\%$ and a duration of $15 \mu s$\cite{bochen2024compilation} (green dots in Fig.~\ref{fig:background}(a)).

However, the collective movement within an AOD lattice must adhere to the following constraints (Fig.\ref{fig:background}(c)):\\
\noindent (1) \emph{Rows and columns must move in tandem and cannot cross.} This means the AOD frame can stretch or contract in two directions, but the relative order of rows and columns must remain fixed~\cite{bluvstein2022quantum}.\\
\noindent (2) \emph{Movement speed must be controlled.} Experiments~\cite{bluvstein2022quantum} show that qubit fidelity is maintained as long as the acceleration does not exceed $a = 2750m\cdot s^{-2}$.

\noindent Notably, NAQC can support multiple independently operating AOD arrays, such as the yellow and green lattices shown in Fig.~\ref{fig:background}(a). Qubit movements in distinct AOD arrays can be performed simultaneously, which enhances parallelism.

\begin{table}[htbp]
\centering
    \begin{tabular}{| c | c | c | c | c |}
        \hline
        & 1Q Gate & CZ Gate & Excitation & Transfer \\
        \hline
        Fidelity & $99.99\%$ & $99.5\%$ & $99.75\%$ & $99.9\%$ \\
        \hline
        Duration & $1\mu s$ & $270ns$ & $270ns$ & $15\mu s$ \\
        \hline
        \hline
        \multicolumn{1}{|c|}{} & \multicolumn{4}{c|}{Qubit Movement} \\
        \hline
        \multicolumn{1}{|c|}{Fidelity} & \multicolumn{4}{c|}{$\sim 100\%$ if $a<2750m\cdot s^{-2}$} \\
        \hline
        \multicolumn{1}{|c|}{Duration} & \multicolumn{4}{c|}{e.g. $100\mu s(200\mu s)$ for $27.5\mu m (110\mu m)$} \\
        \hline
    \end{tabular}
\centering
 \vspace{3pt}
\caption{Parameters on the fidelity and duration of operations on NAQC.}
\vspace{-6pt}
\label{tab:NA hardware parameter}
\end{table}


\noindent\textbf{Zoned Architecture.} The ZA has been physically demonstrated~\cite{bluvstein2024logical}, dividing the computational space into distinct zones for specific tasks. Although originally designed for logical qubits in quantum error correction codes~\cite{knill1997theory}, ZA offers new opportunities for optimizing near-term applications with bare qubits. For example, a storage zone can be spatially separated from the computation zone (at least $20\mu m$ away in~\cite{bluvstein2024logical}), as shown in Fig.~\ref{fig:background}(a). Qubits held in the storage zone are well-preserved and unaffected by Rydberg excitation, avoiding both decoherence and excitation errors. Qubits can be shuttled between these zones as needed.

For more details of hardware features, please refer to~\cite{bluvstein2024logical, schmid2024computational, wurtz2023aquila}. We summarize the hardware parameters into Table~\ref{tab:NA hardware parameter}.

\subsection{Fidelity Analysis}\label{subsec:fidelity analysis}
This subsection presents a comprehensive fidelity analysis that informs our optimization objectives. 

The output fidelity can be decomposed into five components: (1) 1Q gates, (2) CZ gates, (3) excitation error, (4) transfer error, and (5) decoherence error. The first two components are computed as $f_1^{g_1}$ and $f_2^{g_2}$, where $f_1=99.99\%$ and $f_2=99.5\%$ are the fidelities of 1Q and CZ gates, respectively (see Table~\ref{tab:NA hardware parameter}), and $g_1$ and $g_2$ are the number of 1Q and CZ gates. Qubits remaining in the computation zone without a CZ gate acting on them will still be excited by the Rydberg laser, and later return to the original state, causing a fidelity reduction. The excitation error is given by $f_{exc}^{\sum_{i=1}^S n_i}$, where $S$ is the total number of Rydberg excitations, $n_i$ is the number of non-interacting qubits during the $i$-th excitation, and $f_{exc} = 99.75\%$. The transfer error is expressed as $f_{trans}^{N_{trans}}$, where $f_{trans}=99.9\%$ and $N_{trans}$ represents the total number of qubit transfers. Qubits also experience decoherence when not involved in gate operation, called the \emph{idle periods} (e.g., during transfer or movement). Let $T_q$ represent the total idle time for qubit $q$, resulting in decoherence error $1 - T_q / T_2$, where $T_2 = 1.5s$~\cite{bluvstein2024logical, bluvstein2022quantum} is the coherence time of neutral atom qubits. However, decoherence can be mitigated by moving qubits to the storage zone, where coherence decay is assumed to be negligible~\cite{bluvstein2024logical}. Combining these factors, the output fidelity is computed as:
\begin{equation}\label{fidelity formula}
f_{output} = f_1^{g_1}\cdot f_2^{g_2}\cdot f_{exc}^{\sum_{i=1}^S n_i}\cdot f_{trans}^{N_{trans}}\cdot \prod_{q}\left(1-\frac{T_q}{T_2}\right)
\end{equation}


In practice, the input benchmark circuits are synthesized into alternating layers of 1Q gates and CZ gate blocks~\cite{wang2024atomique, tan2024compiling,bochen2024compilation}. Since the 1Q gate layers can be executed conveniently (see Sec.~\ref{subsec: NA hardware cap}), compiler optimization typically focuses on the CZ gate blocks, and the 1Q term in equation (\ref{fidelity formula}) is often omitted in fidelity comparisons.

\section{Motivation}\label{sec:Motivate}
In this section, we analyze the limitations of existing work (Sec.\ref{subsec:limitation}), with a particular focus on the current leading approach Enola~\cite{bochen2024compilation}. We provide concrete examples to demonstrate how its limitations arise from addressing various aspects of the problem in isolation, as mentioned in Sec.\ref{sec:intro}. Building on this analysis, we point out the key motivations behind the three core components of our solution.


\subsection{Limitations of Existing NAQC Compilers}\label{subsec:limitation}
The compiler Enola~\cite{bochen2024compilation} currently offers the best performance. Previous work~\cite{wang2024atomique, wang2023q} introduces additional two-qubit gates for qubit interaction, which significantly reduces fidelity. In contrast, Enola introduces no extra gates beyond those in the input program. Solver-based methods~\cite{tan2022qubit, tan2024compiling} also avoid additional gates but limit flexibility in layout transitions, resulting in more stages, more Rydberg excitations, and higher excitation error. Enola optimizes the number of stages and minimizing excitation error, while also using efficient heuristics to address scalability. 

However, Enola has two major drawbacks that limit its performance: (1) \emph{Suboptimal movement scheme} and (2) \emph{Challenges with storage zone integration}. We provide examples to illustrate each of these issues.



\begin{figure}[!ht]
    \centering
    \includegraphics[width=0.47\textwidth]{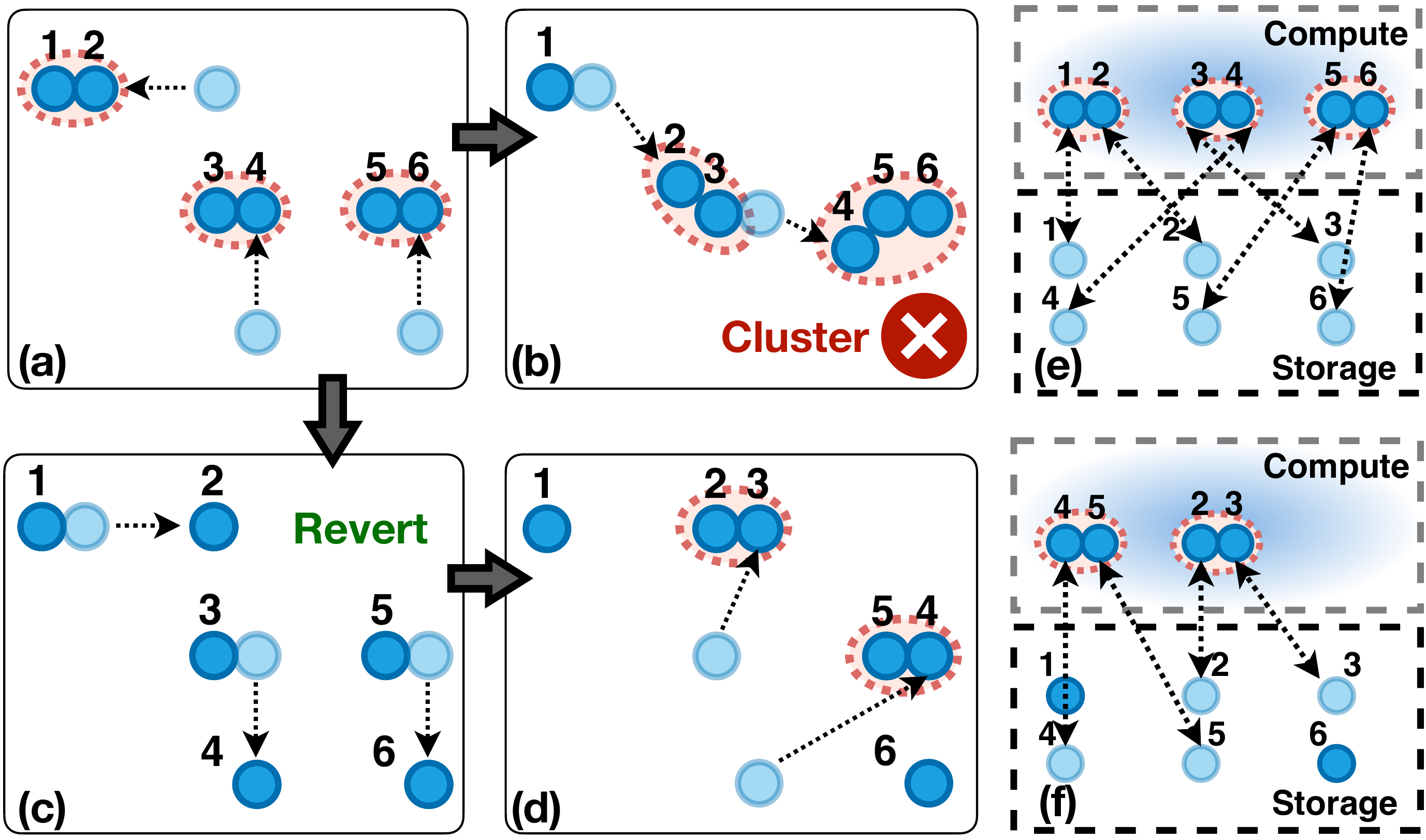}
    \vspace{-6pt}
    \caption{(a)-(d) Qubit clustering issue in Enola. (e)-(f) Challenges with Enola’s integration of the storage zone.}
    \Description{Constraints}
    \label{fig:motivation} 
\end{figure}

\vspace{0.3em}
\noindent\textbf{Example 1. Suboptimal Movement Scheme.} Enola’s movement scheme reverts to the initial layout before transitioning to the next stage. The reason is that a \emph{direct} layout transition leads to unwanted qubit clustering. For example, in Fig.~\ref{fig:motivation}(a), qubit pairs $(q_1,q_2)$, $(q_3,q_4)$, and $(q_5,q_6)$ are positioned close for CZ execution. In the next stage, CZ gates are needed on pairs $(q_2,q_3)$ and $(q_4,q_5)$. Enola would move $q_2$ to $q_3$ and $q_4$ to $q_5$, but this creates a cluster of $q_4,q_5,q_6$, preventing the desired CZ gate on $(q_4,q_5)$, as shown in Fig.~\ref{fig:motivation}(b). To avoid this clustering, Enola reverts to the initial layout, spatially separating the qubits (Fig.\ref{fig:motivation}(c)). From this layout, interacting qubits can then be brought together without causing clustering issues (Fig.\ref{fig:motivation}(d)). However, repeatedly returning to the initial layout introduces significant movement overhead, which could be minimized by directly transitioning between desired layouts for parallel CZ execution.

\vspace{0.3em}
\noindent\textbf{Example 2. Challenges with Storage Zone Integration.} Enola’s framework \emph{does not} incorporate a storage zone due to the recent emergence of the ZA and is confined solely to the computation zone. Its movement scheme, however, limits the efficient integration of a storage zone. Enola requires reverting to the initial layout between stages. Therefore, to integrate a storage zone, this initial layout would need to be entirely placed in the storage zone to avoid excitation errors, as shown in Fig.\ref{fig:motivation}(e). For each stage, interacting qubits would need to shuttle back and forth between the storage and computation zones to return to the initial layout. For instance, to execute the two stages shown in Fig.\ref{fig:motivation}(a)(d), Enola would need to move the qubits as depicted in Fig.~\ref{fig:motivation}(e)(f), resulting in significant inter-zone movement overhead.

\vspace{0.3em}
\noindent\textbf{Analysis.} The core reason behind Enola’s limitations lies in its decomposition of the entire problem into three sub-problems, addressing \emph{gate scheduling}, \emph{qubit allocation}, and \emph{qubit movement} in isolation. While this approach yields optimal solutions for each sub-problem, the overall solution is suboptimal because it overlooks opportunities for deeper optimization that arise from the synergy between these subproblems. For example, Enola first optimizes qubit allocation to obtain an initial layout. However, the subsequent optimization of qubit movement is constrained by this fixed layout, hindering direct transitions between layouts for CZ execution (Example 1). This leads to a layout that is effectively semi-static, underutilizing the dynamic potential of NAQC and further preventing the efficient integration of the storage zone (Example 2). 

This insight reveals a key motivation for our approach: rather than treating these aspects in isolation, we seek to recognize and exploit their interdependencies, unlocking new optimization possibilities that a segmented approach fails to capture. Building on this, we developed three key components that form the core of our solution, which we will introduce in detail in Sec.~\ref{sec:stage scheduler}, Sec.~\ref{sec:conti router}, and Sec.~\ref{sec:Move scheduler}.

\section{Stage Scheduler}\label{sec:stage scheduler}
In this section, we focus on minimizing the number of Rydberg stages and optimizing the interplay between gate scheduling and zoned architecture to reduce decoherence errors. The \emph{Stage Scheduler} first partitions the program circuit into stages. Within each stage, CZ gates can be executed within a single Rydberg excitation. It then determines the execution order of these stages to minimize inter-zone communication between stages. 

\begin{figure*}[!ht]
    \centering
    \includegraphics[width=1\textwidth]{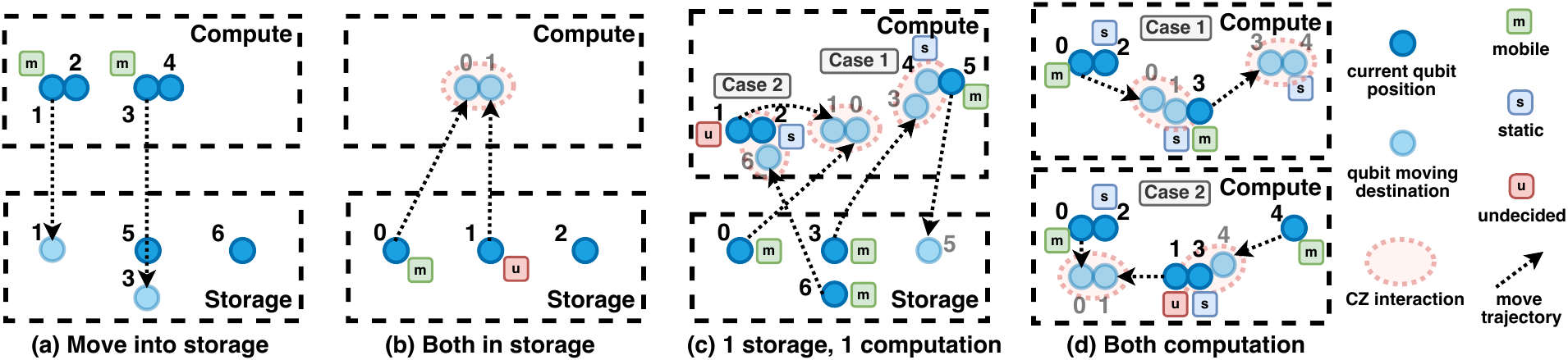}
    \caption{
    Illustration of four qubit movement scenarios: (a) Two qubits in the computation zone move to the storage zone in the next stage as they do not interact with other qubits. (b) Two qubits in the storage zone move to the computation zone for a CZ interaction. (c) We consider two cases in which one qubit is in the computation zone and the other is in the storage zone, with both moving to the computation zone for interaction. In this figure, the order of label assignment (\textbf{mobile}, \textbf{static}, \textbf{undecided}) for \emph{Case 1} is \{\textbf{6}, \textbf{2}, \textbf{0}, \textbf{1}\}, and the order for \emph{Case 2} is \{\textbf{5}, \textbf{3}, \textbf{4}\}. (d) Two cases involving qubits both located in the computation zone that require interaction. In \emph{Case 1} shown in this figure, qubit \textbf{2} has already been assigned the \textbf{static} label. In the \emph{Case 2}, qubits \textbf{2} and \textbf{4} have both been labeled as \textbf{static}. The order of label assignment for \emph{Case 1} is \{\textbf{3}, \textbf{4}, \textbf{0}, \textbf{1}\}, and the order for \emph{Case 2} is \{\textbf{3}, \textbf{0}, \textbf{1}\}.
    }
    \Description{Technical}
    \label{fig:continuous router}
\end{figure*}

\subsection{Stage Partition}
In this step, we first divide the program circuit into dependent CZ blocks each consisting of commutable CZ gates. We then partition each CZ gate block 
into stages, which are groups of CZ gates acting on disjoint qubits, allowing them to be executed in parallel. 


We use an optimized edge-coloring algorithm for stage partitioning, as shown in Algorithm \ref{alg:optimized_coloring}. Given an input list of CZ gates, each gate is assigned a color. Within a stage, CZ gates acting on overlapping qubits must be executed in separate stages, so they should be assigned different colors. For each CZ gate, if it does not share any interacting qubits with the currently colored stages, it is assigned the same color and grouped into that stage. If it shares interacting qubits with all existing stages, it is assigned a new color and placed in a new stage. Once all CZ gates have been processed, the partitioning into stages is complete.

\begin{algorithm}
\caption{Stage Partition Algorithm}
\label{alg:optimized_coloring}
\footnotesize
\KwData{CZ\_Graph(CZ Interaction Graph)}
\KwResult{Stages(Partitioned Stages)}

\SetKwFunction{SortVerticesByDegree}{sortVerticesByDegree}
\SetKwFunction{AssignColor}{AssignColor}
\SetKwFunction{OptimizedColoring}{OptimizedColoring}
\SetKwFunction{CollectStages}{CollectStages}

\SetKwProg{Fn}{Function}{:}{}
\Fn{\AssignColor{$vertex$, $color$, $available$}}{
    $available \leftarrow \text{True array of size } n$;\\
    \For{each $u \in \text{CZ\_graph.adjacents}(vertex)$}{
        \If{$color[u] \neq -1$}{
            $available[color[u]] \leftarrow \text{False}$;\\
        }
    }
    \For{each $c \gets 0 \text{ to } n$}{
        \If{$available[c]$}{
            $color[vertex] \leftarrow c$;\\
            \textbf{break};\\
        }
    }
}

\SetKwProg{Fn}{Function}{:}{}
\Fn{\OptimizedColoring{CZ\_Graph}}{
    $n \gets \text{number of vertices in CZ\_Graph}$;\\
    $color \leftarrow \text{array of size } n \text{ initialized to } -1$;\\
    \textit{\textbf{// Sort vertices in descending order by degree}}\\
    $sortedVertices \leftarrow \SortVerticesByDegree(CZ\_graph)$;\\
    
    \For{each $v \in sortedVertices$}{
        \AssignColor{$v$, $color$, $available$};\\
    }
    \textit{\textbf{// Collect stages according to the colored graph.}}\\
    $Stages \leftarrow \CollectStages(color)$;\\
    \Return $Stages$;\\
}
\end{algorithm}

\subsection{Stage Scheduling}
In this step, we schedule the stages generated by a CZ block to minimize qubit interchange between zones, thereby reducing movement overhead due to the integration of zoned architecture. Since the CZ block consists of commutable gates, the execution sequence of its generated stages can be freely rearranged. First, an initial layout is placed entirely in the storage zone. Since the layout will change continuously during computation without returning to this initial configuration, its role is less significant compared to previous works like~\cite{bochen2024compilation}. For convenience, we adopt the initial layout from that work.

We select the first stage to be the one with the fewest interacting qubits, allowing as many qubits as possible to remain in the storage zone, thus reducing decoherence error. Next, we greedily select the subsequent stage to be the one that differs the least in the set of interacting qubits from the current stage. Let the sets of interacting qubits for the current stage \( S_i \) and the next stage \( S_{i + 1} \) be denoted as \( Q_i \) and \( Q_{i + 1} \), respectively. We quantify the difference between the two stages by 

\[
|Q_i \setminus Q_{i + 1}| + \alpha |Q_{i + 1} \setminus Q_i|,
\]
where we assign a lower weight \( \alpha < 1 \) to the term $|Q_{i + 1} \setminus Q_i|$. This preference reflects our desire for qubits to move into storage rather than out of it, as qubits in the storage zone experience negligible decoherence errors.

\section{Continuous Router}\label{sec:conti router}

This section introduces the \emph{continuous router} in our solution. 
Compared to earlier compilers that revert to their initial layout after each Rydberg excitation to prevent clustering, we utilize a more efficient algorithm that allows qubits to transition directly into the layout for the next stage's CZ execution. We assume that the stage scheduler has given an ordered list of CZ stages (introduced in Sec.~\ref{sec:stage scheduler}). The continuous router consists of two steps: \textbf{(1) Single Qubit (1Q) Movement Decision}, which determines the 1Q movements required to facilitate CZ gates and inter-zone communication in the next stage. \textbf{(2) 
 Coll-Move Grouping}, which groups the 1Q movements from the previous step into Coll-Moves while adhering to movement constraints.

\subsection{Basic Set-ups}
Before we start, we introduce some notations and basic set-ups for describing our solution. We assume the \emph{qubit sites} are on a 2D grid and denote them by the coordinates $(x,y)$. We set the minimal spatial distance between sites as $15\mu m$ according to~\cite{bluvstein2024logical}.
We assume a qubit can only stay in a site when it's not moved and specify their locations by the coordinates of sites. A site can either hold two interacting qubits, or one non-interacting qubit, or can be empty. We assume the storage zone and computation zone are spatially separated by $30\mu m$~\cite{bluvstein2024logical}. 


\subsection{Single Qubit Movement Decision}
This subsection illustrates how we decide 1Q movements needed for the next stage given the current qubit layout. These 1Q movements should enable all the intended CZ gates, qubit interchange between computation and storage zones, and not induce unwanted clustering of qubits. 

We characterize 
the 1Q movements by assigning each qubit $q$ a target site location $(x^q_{target},y^{q}_{target})$. This unified representation simplifies the problem, as specifying the site coordinates of the computation and storage zones eliminates the distinction between inter- and intra-zone movements. As a result, the same representation can be applied to both, streamlining the process.


The determination of 1Q movements follows three steps.

\vspace{3pt}\noindent\textbf{Step 1. Determine 1Q movements for non-interacting qubits. }The non-interacting qubits in the next stage will be labeled as \emph{mobile} and moved into storage. We move each of them vertically down to the closest empty site in storage. We determine these 1Q moves following the descending order of y-coordinates of qubits, so that qubits farther from the storage zone can choose their sites first, which decreases the total movement distance. For example, the non-interacting qubit \textbf{1} and \textbf{3} in Fig.~\ref{fig:continuous router}(a) is moved to its nearest available site in the storage zone.

\vspace{3pt}\noindent\textbf{Step 2. Assign labels to interacting qubits.} 
We assign each qubit a label: \textbf{static}, \textbf{mobile}, or \textbf{undecided}. In the layout rearrangement, \textbf{static} qubits remain in their current positions, waiting for other qubits to move in for interaction. \textbf{Mobile} qubits, on the other hand, will move to other sites (either interaction sites or storage). We designate certain qubits as \textbf{undecided} for two reasons: either their current positions already contain a static qubit, necessitating their movement to avoid clustering, or both qubits intended for interaction are located in the storage zone, so their interaction site needs to be determined.

For each CZ gate $(q_i,q_j)$ in the next stage, there are only four possibilities for the current locations of $q_i,q_j$:

\textbf{(1) $q_i,q_j$ are both in storage.}
To conduct a CZ gate, $q_i$ and $q_j$ need to be moved to the same site in computation zone. This site will be decided later in Step 3 considering the sites of other qubits. As a result, we set one of the qubit $q_j$ as undecided, and set the other qubit $q_i$ as mobile, with its site decided by that of $q_j$ ($q_i\rightarrow q_j$) in Step 3.
This is illustrated in Fig.~\ref{fig:continuous router}(b) by qubit \textbf{0} and \textbf{1}.

\textbf{(2) $q_i$ is in storage, $q_j$ is in computation zone.} We first set $q_i$ as mobile,
since it has to move out from storage anyway. 
We then set \( q_j \) as static or undecided based on whether its site already contains a static qubit, subsequently determining the moving destination of \( q_i \) accordingly. \textbf{Case 1. } If $q_j$'s site has no other static qubits, we can set $q_j$ as static and determine the move of $q_i$ ($q_i\rightarrow q_j$). For example, in \emph{Case 1} of Fig.~\ref{fig:continuous router}(c), the site of qubit \textbf{4} has no other static qubits because the other qubit \textbf{5} in the site will be moved to the storage. Therefore, qubit \textbf{4} can be set to static, determining the move \textbf{3} $\rightarrow$ \textbf{4}. \textbf{Case 2. }If $q_j$'s site already contains a static qubit, then we set $q_j$ as undecided due to the potential qubit clustering. For example, in \emph{Case 2} of Fig.~\ref{fig:continuous router}(b), there is a static qubit \textbf{2} at the same site with qubit \textbf{1}, so qubit \textbf{1} has to be set as undecided. The move \textbf{0} $\rightarrow$ \textbf{1} will be completely decided once the target location of qubit \textbf{1} is decided in Step 3. 

\textbf{(3) $q_j$ is in storage, $q_i$ is in computation zone.} This case is symmetric to (2) by interchanging the role of $q_i$ and $q_j$, hence we omit the discussion.

\textbf{(4) $q_i$ and $q_j$ are in the computation zone.}
In this case, either $q_i$ and $q_j$ has to move. We randomly set one of them as mobile and then set the other qubit as static or undecided based on whether there is a static qubit in its site.
As illustrated in  Fig.~\ref{fig:continuous router}(d), if qubit \textbf{1}'s site has no other static qubits, we set qubit \textbf{1} as static and decide the move \textbf{0}$\rightarrow$\textbf{1} (\emph{Case 1} in Fig.~\ref{fig:continuous router}(d)); otherwise if qubit \textbf{1}'s site has static qubits, we set qubit \textbf{1} as undecided and determines the move \textbf{0} $\rightarrow$ \textbf{1} later (\emph{Case 2} in Fig.~\ref{fig:continuous router}(d)).


\vspace{3pt}\noindent\textbf{Step 3. Determine the target site for ``undecided'' qubits.} Let’s recall that in Step 2, we designated some qubits as undecided, awaiting a new location for them to move. We search around its current location to find the nearest empty site in the computation zone and set it as the target location of this undecided qubit, and its associated interacting qubit will move to this site. Fig.~\ref{fig:continuous router}(c) \emph{Case 2} and Fig.~\ref{fig:continuous router}(d) \emph{Case 2} give two such examples.\\

After the above three steps, the 1Q movements for each qubit have been precisely determined. Next, we group them into Coll-Moves that can be executed within an AOD array, aiming to minimize the total movement time for layout rearrangement.
\begin{figure}[h!]
    \vspace{-5pt}
    \includegraphics[width=0.98\linewidth]{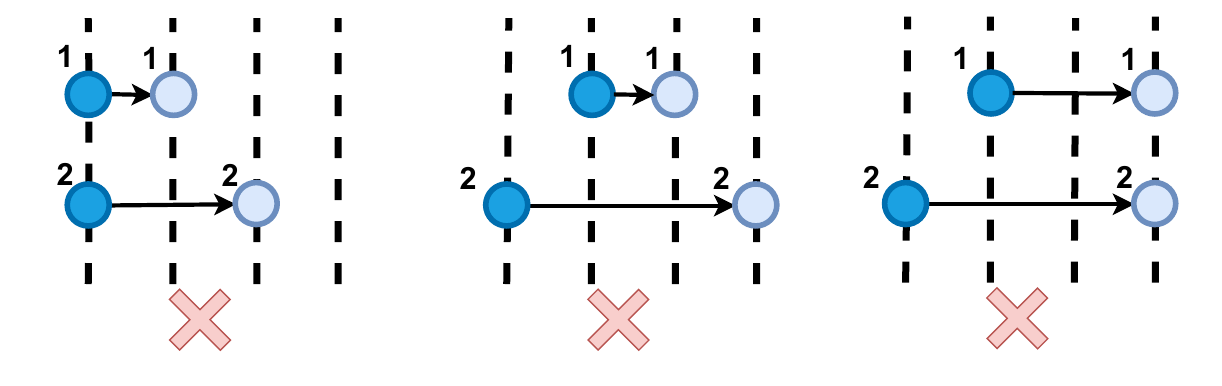}
    
    \textbf{$x^1_{start} = x^2_{start}$}\hspace{27pt}
    \textbf{$x^1_{start} > x^2_{start}$}\hspace{19pt}
    \textbf{$x^1_{start} > x^2_{start}$}\\
    \textbf{$x^1_{end} \neq x^2_{end}$}\hspace{36pt}
    \textbf{$x^1_{end} < x^2_{end}$}\hspace{28pt}
    \textbf{$x^1_{end} = x^2_{end}$}\\
    
    \caption{Movement conflicts on x-coordinate.}
    \vspace{-5pt}
    \label{fig: x conflicts}
\end{figure}

\subsection{Collective Movement Grouping}
Since the movement constraints within an AOD (Sec.~\ref{subsec: NA hardware cap}) may not allow all the 1Q movements to be conducted simultaneously, we group them into collective moves (Coll-Moves) with two optimization goals: (1) minimizing the total number of Coll-Moves, (2) minimizing the maximal movement distance for each Coll-Moves, since it determines the movement time. Both objectives aim to reduce execution time and minimize decoherence errors. To achieve these goals, we introduce a \emph{distance-aware} grouping method. The key idea is to greedily grouping the 1Q movements following the ascending order of movement distance.

Before we describe the algorithm, we define the notion of two 1Q movements \emph{conflicting} with each other, since it serves as the criterion for grouping to Coll-Moves: the 1Q moves within a Coll-Move should not conflict. A conflict happens when the order of $x$- or $y$-coordinate of two moving qubits changes after the movement. The rigorous definition is as follows. Assuming that there are two moves:
\begin{align*}
    &m_1 = (x^1_{start},y^{1}_{start}) \rightarrow(x^1_{end},y^{1}_{end})\\
    &m_2 = (x^2_{start},y^{2}_{start})\rightarrow(x^2_{end},y^{2}_{end})
\end{align*}
where the coordinates represent site locations. We say $m_1$ and $m_2$ conflict on $x$-coordinate if
$x^1_{start} \leq x^2_{start}$ but $x^1_{end} > x^2_{end}$, or $x^1_{start} \geq x^2_{start}$ but $x^1_{end} < x^2_{end}$, as illustrated in Fig. \ref{fig: x conflicts}. Similarly, we define the conflicts on $y$-coordinate for two 1Q moves. Finally, we say $m_1$ and $m_2$ if they conflict either on $x$- or $y$-coordinate. 

We first sort the 1Q movements in the ascending order of movement distance: $m_1,m_2,\cdots$. Assuming that we have assigned $m_1$ to $m_n$ into Coll-Moves groups $G_1,\cdots, G_k$ and we want to assign $m_{n+1}$ to a group. We check if the 1Q movement $m_{n+1}$ conflicts with any of $G_i$. If there is no conflict, we assign $m_{n+1}$ to $G_i$, otherwise we check the conflict condition for the next group. If $m_{n+1}$ cannot be assigned to any group, then it's assigned to a new Coll-Moves group $G_{k+1}$. Notably, this method tends to group movements with similar distance together, potentially reduces the total movement time. This is because the movement time of a Coll-Move is determined by the longest-distance 1Q movement in it, so a grouping with balanced distance can suppress the movement time.

\section{Coll-Moves Scheduler}\label{sec:Move scheduler}
In this section, we optimize the execution order of Coll-Moves. Additionally, we utilize multiple AOD arrays for parallel processing, effectively leveraging hardware resources to enhance fidelity and suppress execution time.
\subsection{Intra-Stage Scheduler}

To minimize decoherence errors brought by the interplay with ZA architecture, we optimize the execution order of Coll-Moves by prioritizing move-in operations to the storage zone while delaying move-out operations. We achieve this by first grouping the Coll-Moves and then scheduling the execution sequence based on the difference between the number of move-in and move-out operations. Specifically, for each Coll-Move group \(G_i\), we denote the number of move-in operations as \({n_{\text{in}}}^i\) and the number of move-out operations as \({n_{\text{out}}}^i\). We sort the Coll-Move groups in descending order of \({n_{\text{in}}}^i - {n_{\text{out}}}^i\), resulting in the final execution sequence \(\{G_1', \cdots, G_k'\}\). This order prioritizes Coll-moves with a greater number of move-in operations, ensuring they are performed earlier. As a result, qubits will stay in the storage zone for longer periods, thereby reducing their exposure to decoherence.

\subsection{Multi-AOD Scheduler}\label{sec:multi-AOD scheduler}
Using multiple AOD arrays can further parallelize the execution of Coll-Moves. In the case of a single AOD, considering the constraints mentioned in Section \ref{subsec: NA hardware cap}, conflicting movements cannot be executed together. However, in the multiple AODs scenario, 1Q movements from different AODs may conflict but they can still be executed in parallel because different AODs operate independently. This enables the distribution of previously conflicting qubit movements across different AODs, allowing for the parallel execution of more qubit movements. Given \(n\) AODs and having scheduled the Coll-Move groups into \(\{G_1', \cdots, G_k'\}\) with corresponding maximum movement durations \(\{t_1', \cdots, t_k'\}\), we divide them into \(m\) parallel groups:

\[
\{G_1', \cdots, G_n'\}, \cdots, \{G_{(m-1)n+1}', \cdots, G_k'\}.
\]
For the \(r\)-th parallel group \(\{G_{(r-1)n+1}', \cdots, G_{rn}'\}\), the execution duration is given by \(t_{\text{transfer}} + \max(t_{(r-1)n+1}', \cdots, t_{rn}')\). This parallelism reduces the total transfer and movement duration, thus suppressing the decoherence error. We point out that the transfer error term in the fidelity formula (\ref{fidelity formula}) is not affected because the number of transfers does not change. 


\begin{table}[h]
    \centering
      \caption{Benchmarks.}
      
    \resizebox{0.48\textwidth}{!}{
        \renewcommand*{\arraystretch}{1}
        \begin{footnotesize}
        \begin{tabular}{|p{1.7cm}|p{1.5cm}|p{2.0cm}|p{1.4cm}|p{1.6cm}|}  \hline
        Name & \#Qubits & Compute Zone Size (${\mu m}^2$) & Inter Zone Size (${\mu m}^2$) & Storage Zone Size (${\mu m}^2$)\\ 
        \hline
        \multirow{6}{*}{QAOA-regular3}  
                     & 30 & 90 x 90 & 90 x 30 & 90 x 180\\ 
        \cline{2-5}  & 40 & 105 x 105 & 105 x 30 & 105 x 210\\
        \cline{2-5}  & 50 & 120 x 120 & 120 x 30 & 120 x 240\\
        \cline{2-5}  & 60 & 120 x 120 & 120 x 30 & 120 x 240\\
        \cline{2-5}  & 80 & 135 x 135 & 135 x 30 & 135 x 270\\
        \cline{2-5}  & 100 & 150 x 150 & 150 x 30 & 150 x 300\\
        \hline
        \multirow{5}{*}{QAOA-regular4}  
                     & 30 & 90 x 90 & 90 x 30 & 90 x 180\\ 
        \cline{2-5}  & 40 & 105 x 105 & 105 x 30 & 105 x 210\\
        \cline{2-5}  & 50 & 120 x 120 & 120 x 30 & 120 x 240\\
        \cline{2-5}  & 60 & 120 x 120 & 120 x 30 & 120 x 240\\
        \cline{2-5}  & 80 & 135 x 135 & 135 x 30 & 135 x 270\\
        \hline
        \multirow{2}{*}{QAOA-random}  
                     & 20 & 75 x 75 & 75 x 30 & 75 x 150\\ 
        \cline{2-5}  & 30 & 90 x 90 & 90 x 30 & 90 x 180\\
        \hline
        \multirow{2}{*}{QFT}  
                     & 18 & 75 x 75 & 75 x 30 & 75 x 150\\ 
        \cline{2-5}  & 29 & 90 x 90 & 90 x 30 & 90 x 180\\ 
        \hline
        \multirow{3}{*}{BV}  
                     & 14 & 60 x 60 & 60 x 30 & 60 x 120\\ 
        \cline{2-5}  & 50 & 120 x 120 & 120 x 30 & 120 x 240\\
        \cline{2-5}  & 70 & 120 x 120 & 120 x 30 & 120 x 240\\
        \hline
        \multirow{2}{*}{VQE}  
                     & 30 & 90 x 90 & 90 x 30 & 90 x 180\\ 
        \cline{2-5}  & 50 & 120 x 120 & 120 x 30 & 120 x 240\\
        \hline
        \multirow{3}{*}{QSIM-rand-0.3}  
                     & 10 & 60 x 60 & 60 x 30 & 60 x 120\\ 
        \cline{2-5}  & 20 & 75 x 75 & 75 x 30 & 75 x 150\\ 
        \cline{2-5}  & 40 & 105 x 105 & 105 x 30 & 105 x 210\\
        \hline

        \end{tabular}
        \end{footnotesize}
    }
    \label{tab:benchmark}
\end{table}

\section{Evaluation}\label{sec:eval}
\begin{table*}[tp]
    \centering
    \caption{The results of our compiler and its relative performance to the baseline. Each benchmark-$n$' corresponds to an $n$-qubit circuit in the circuit model.}
   \resizebox{\textwidth}{!}{
        \renewcommand*{\arraystretch}{1.3}
        \begin{large}
            
            \begin{tabular}{|p{3.2cm}|| p{1.6cm}|p{2.3cm}|p{2.3cm}|p{1.7cm}||p{2.1cm}|p{2.3cm}|p{2.3cm}|p{2.1cm}||p{2.3cm}|p{2.1cm}|p{2.3cm}|}
        
        \hline
        Benchmark - \#Qubit & Enola Fidelity& Our Fidelity (non-storage) & Our Fidelity (with-storage) & Fidelity \ \ Improv.  & Enola $T_{exe}(\mu s)$ & Our $T_{exe}(\mu s)$ (non-storage) & Our $T_{exe}(\mu s)$ (with-storage) & $T_{exe}$ Improv. & Enola $T_{comp}(s)$ & Our $T_{comp}(s)$ & $T_{comp}$ Improv.\\
        \hline
        \hline

        QAOA-regular3-30 & 0.48 & 0.64 & 0.68 & 1.41 & 13,198.04 & 4,680.72 & 6,116.19 & 2.82 & 128.32 & 41.33 & 3.10\\
        \hline
        QAOA-regular3-40 & 0.34 & 0.53 & 0.57 & 1.67 & 17,249.38 & 5,601.12 & 8,998.75 & 3.08 & 144.70 & 41.50 & 3.49\\
        \hline
        QAOA-regular3-50 & 0.23 & 0.43 & 0.49 & 2.12 & 21,087.88 & 7,135.26 & 9,582.99 & 2.96 & 142.30 & 41.49 & 3.43\\
        \hline
        QAOA-regular3-60 & 0.14 & 0.35 & 0.39 & 2.70 & 25,449.73 & 8,134.16 & 12,440.46 & 3.13 & 140.64 & 44.62 & 3.15\\
        \hline
        QAOA-regular3-80 & 0.05 & 0.22 & 0.24 & 4.90 & 33,553.14 & 10,490.10 & 17,746.76 & 3.2 & 145.91 & 45.38 & 3.22\\
        \hline
        QAOA-regular3-100 & 0.01 & 0.10 & 0.14 & 12.82 & 44,038.42 & 16,122.96 & 21,710.11 & 2.73 & 167.22 & 45.64 & 3.66\\
        \hline
        QAOA-regular4-30 & 0.40 & 0.56 & 0.56 & 1.42 & 16,450.23 & 6,056.05 & 12,127.03 & 2.72 & 256.88 & 65.33 & 3.93\\
        \hline
        QAOA-regular4-40 & 0.24 & 0.45 & 0.42 & 1.72 & 23,365.45 & 7,394.03 & 17,608.55 & 3.16 & 266.53 & 66.07 & 4.03\\
        \hline
        QAOA-regular4-50 & 0.14 & 0.34 & 0.31 & 2.27 & 30,079.41 & 9,928.27 & 20,013.50 & 3.03 & 253.94 & 63.34 & 4.01\\
        \hline
        QAOA-regular4-60 & 0.07 & 0.26 & 0.23 & 3.22 & 36,332.16 & 11,306.93 & 22,594.20 & 3.21 & 278.18 & 68.89 & 4.04\\
        \hline
        QAOA-regular4-80 & 0.01 & 0.10 & 0.09 & 6.06 & 49,182.73 & 19,631.36 & 32,934.94 & 2.51 & 291.68 & 72.17 & 4.04\\
        \hline
        QAOA-random-20 & 0.23 & 0.39 & 0.47 & 2.02 & 32,768.58 & 11,782.99 & 16,845.33 & 2.78 & 960.37 & 136.03 & 7.06\\
        \hline
        QAOA-random-30 & 0.03 & 0.11 & 0.16 & 5.85 & 68,113.52 & 25,391.69 & 38,051.69 & 2.68 & 1791.66 & 193.28 & 9.27\\
        \hline
        QFT-18 & $8.95 \times 10^{-4}$  & $4.87 \times 10^{-3}$ & 0.05 & 60.30 & 108,173.62 & 36,810.15 & 107,637.68 & 2.94 & 10917.80 & 347.47 & 31.42\\
        \hline
        QFT-29 & $7.12 \times 10^{-9}$ & $9.99 \times 10^{-7}$ & $5.78 \times 10^{-4}$ & 81,151.50 & 239,150.00 & 89,670.26 & 237,315.37 & 2.67 & 24116.00 & 511.97 & 47.10\\
        \hline
        BV-14 & 0.57 & 0.60 & 0.91 & 1.58 & 5,583.98 & 3,034.20 & 5,282.11 & 1.84 & 669.48 & 28.79 & 23.26\\
        \hline
        BV-50 & 0.04 & 0.05 & 0.84 & 20.20 & 10,118.96 & 5,631.26 & 9,255.85 & 1.8 & 1710.91 & 17.95 & 95.32\\
        \hline
        BV-70 & $6.92 \times 10^{-4}$ & $1.05 \times 10^{-3}$ & 0.75 & 1,090.36 & 17,620.11 & 10,277.27 & 15,942.37 & 1.71 & 4334.5 & 20.30 & 213.55\\
        \hline
        VQE-30 & 0.71 & 0.81 & 0.79 & 1.12 & 5,436.18 & 1,688.03 & 2,981.71 & 3.22 & 57.62 & 29.68 & 1.94\\
        \hline
        VQE-50 & 0.48 & 0.67 & 0.63 & 1.32 & 10,196.50 & 2,946.26 & 5,354.37 & 3.46 & 56.58 & 29.86 & 1.89\\
        \hline
        QSIM-rand-10 & 0.51 & 0.60 & 0.74 & 1.45 & 13,353.05 & 4,886.36 & 9,713.39 & 2.73 & 760.19 & 76.01 & 10.00\\
        \hline
        QSIM-rand-20 & 0.05 & 0.08 & 0.42 & 9.02 & 37,796.35 & 16,636.02 & 35,550.68 & 2.27 & 5740.76 & 107.03 & 53.64\\
        \hline
        QSIM-rand-40 & $3.94 \times 10^{-6}$ &  $2.39 \times 10^{-5}$ & 0.14 & 35,519.88 & 93,062.71 & 45,424.55 & 89,418.81 & 2.05 & 8283.45 & 127.95 & 64.74\\
        \hline
    \end{tabular}
    \end{large}
    }
    \label{tab:main_table}
\end{table*}

\subsection{Experiment Setup}

\vspace{0.3em}
\noindent \textbf{Hardware setting.}
As outlined in Table \ref{tab:NA hardware parameter}, our hardware configuration follows the latest experimental data. We set the distance between adjacent qubits as \(15 \, \mu m\), and a distance of \(30 \, \mu m\) between the computation zone and the storage zone. 


\vspace{0.3em}
\noindent \textbf{Metrics.} We evaluate the compiler's performance using three metrics. The first metric is \emph{Fidelity}, referring to the overall circuit fidelity, as described in detail in Sec.~\ref{sec:background}. The second metric is the \emph{Execution time}, denoted as \emph{$T_{exe}$}, which accounts for the total time needed for executing single-qubit and two-qubit gates, as well as qubit transfer and movement. The third metric is the \emph{Compilation time}, denoted as $T_{comp}$, which represents the duration taken to transform the high-level quantum program into a low-level implementation suitable for NAQC, including optimization and scheduling processes.

\vspace{0.3em}
\noindent \textbf{Baselines.} 
We primarily focus on comparing PowerMove with Enola~\cite{bochen2024compilation}, as it currently offers the best performance. Specifically, Enola demonstrates a two-qubit fidelity that is 779 times higher than Atomique~\cite{wang2024atomique} and 5806 times higher than Q-Pilot~\cite{wang2023q}, while also mitigating the scalability challenges present in other approaches~\cite{tan2022qubit, tan2024compiling}. Our evaluation supports Enola’s claims regarding these previous works, so we primarily focus on the comparison with Enola for a more concise and relevant analysis.


\vspace{0.3em}
\noindent \textbf{Benchmarks.}
We evaluate performance using a variety of benchmark programs, including Quantum Approximate Optimization Algorithm (QAOA), Quantum Simulation (QSim), Quantum Fourier Transform (QFT), Bernstein-Vazirani (BV) algorithm, and Variational Quantum Eigensolver (VQE). For QAOA, we use two circuit types: one with randomly placed ZZ gates between qubit pairs (50\% probability), and another based on regular graphs, where ZZ gates apply only to qubits connected by graph edges. QSim circuits are randomly generated with a 0.3 probability for a non-identity Pauli operator on each qubit, with ten Pauli strings per circuit. BV circuits use randomly generated secret strings, with an even distribution of 0s and 1s. For VQE, we follow the standard full-entanglement ansatz.

For an \(n\)-qubit program, our default configuration features a \(\lceil\sqrt{n}\rceil \times \lceil\sqrt{n}\rceil\) qubit grid and employs a single AOD array. The storage zone is structured as a \(\lceil\sqrt{n}\rceil \times 2\lceil\sqrt{n}\rceil\) qubit grid. The overall hardware configuration is derived by scaling this grid based on the physical qubit spacing: the compute zone measures \(15 \lceil\sqrt{n}\rceil \times 15  \lceil\sqrt{n}\rceil \, {\mu m}^2\), the inter-zone size is \(15 \lceil\sqrt{n}\rceil\times 30  \, {\mu m}^2\), and the storage zone is \(15 \lceil\sqrt{n}\rceil \times 30 \lceil\sqrt{n}\rceil \, {\mu m}^2\). In Table \ref{tab:benchmark}, we present the benchmarks along with the number of qubits in their circuit representation and the corresponding hardware configuration.


\begin{figure*}[tp]
    \includegraphics[width=0.8\linewidth]{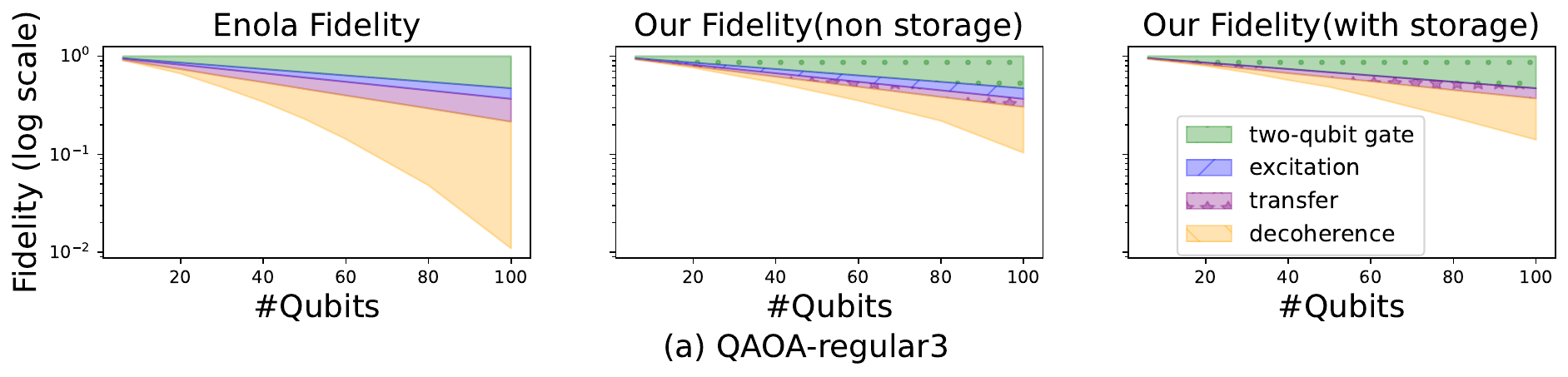}

    \includegraphics[width=0.8\linewidth]{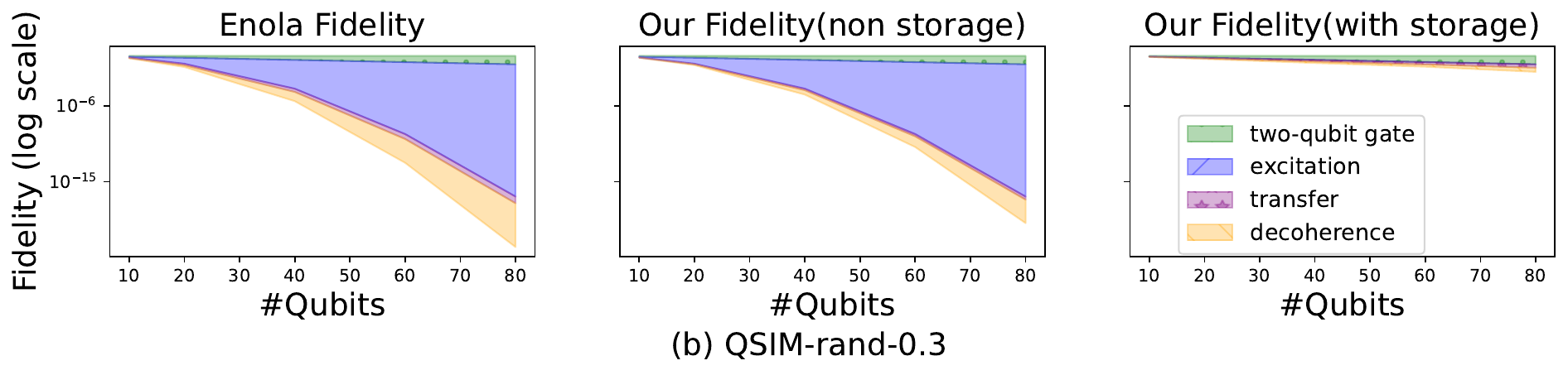}

    \includegraphics[width=0.8\linewidth]{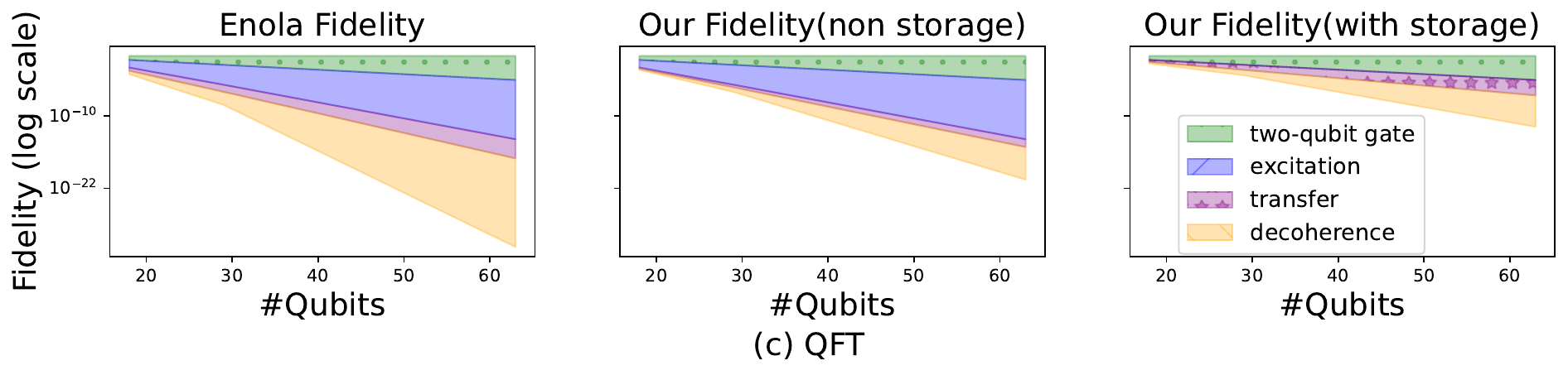}

    \includegraphics[width=0.8\linewidth]{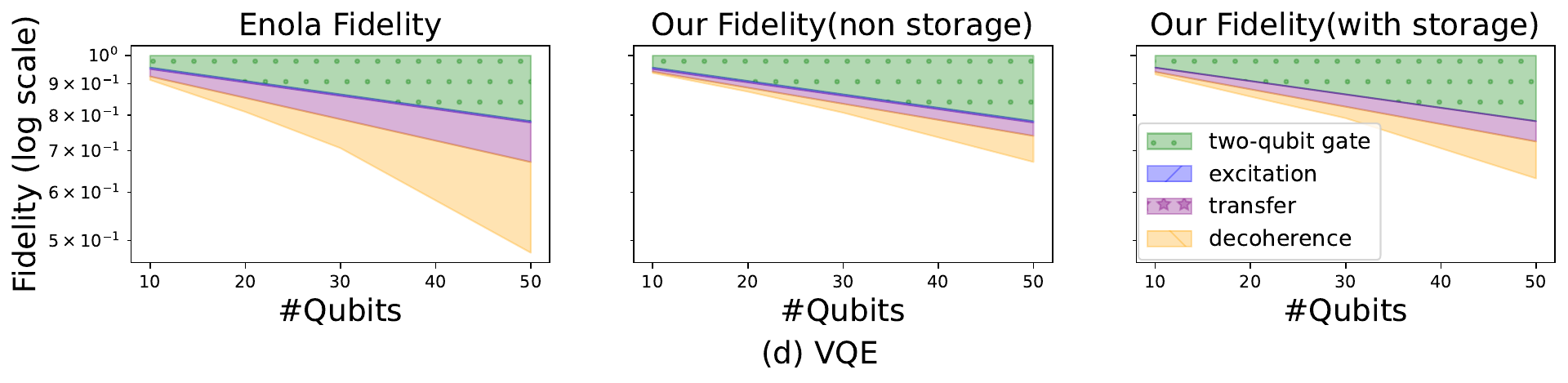}

    \includegraphics[width=0.8\linewidth]{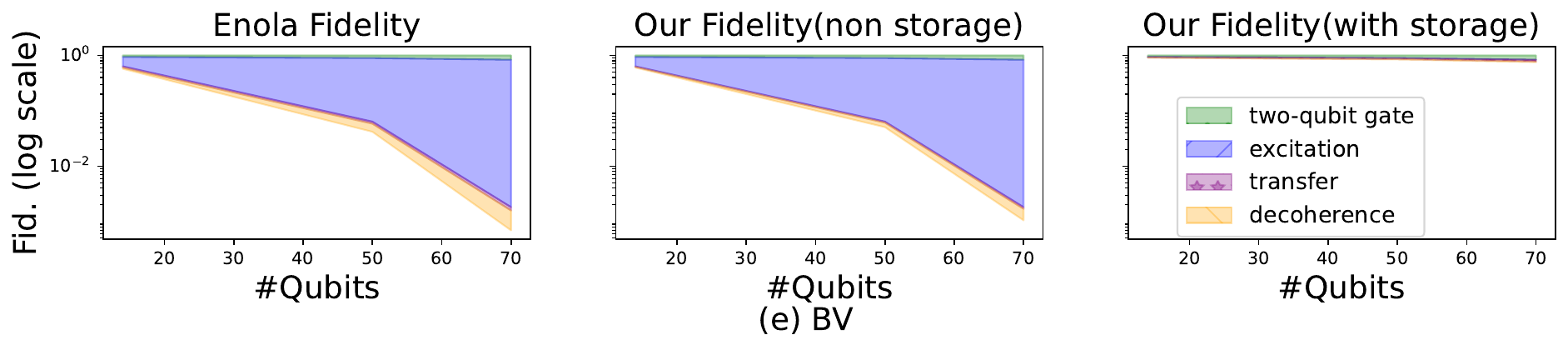}
    \caption{Effects of the continuous router and the introduction of zoned architecture on QAOA-regular3, QSIM-rand-0.3, QFT, VQE, and BV benchmark circuits. This study evaluates various numbers of qubits and focuses on four key components of overall circuit fidelity.}
    \label{fig:fidelity ablation}
\end{figure*}

\subsection{Main Results}\label{subsection: main results}



In this subsection, we compare our compiler with Enola on fidelity and execution time across two scenarios: the \emph{non-storage} case, where only our continuous router is applied, and the \emph{with-storage} case, which also incorporates the other two components regrading the ZA for enhanced performance. We also evaluate the compilation time for both scenarios and report the their average.

\vspace{0.3em}
\noindent \textbf{Overall Performance.}
 As shown in the \emph{Fidelity Improv.} and \emph{$T_{exe}$ Improv.} columns of Table~\ref{tab:main_table}, our framework consistently outperforms the Enola framework, which struggles to achieve reasonable fidelity for large-scale problems. In contrast, our approach enables large-scale programs to maintain high fidelity, particularly in the QSIM-rand and BV benchmarks. For instance, in the 70-qubit BV case, Enola reports a low fidelity of $6.92 \times 10^{-4}$, whereas our method achieves a fidelity of $0.75$, marking a dramatic improvement. Notably, the fidelity improvements increase significantly with the number of qubits, highlighting our framework’s ability to handle much larger programs while ensuring high accuracy—a crucial advantage in the NISQ era. Additionally, the execution time of compiled programs is accelerated by 1.71x to 3.46x. Moreover, we sustain a consistent reduction in execution time as the qubit count increases. These results clearly demonstrate the superior performance of our approach in both fidelity and scalability compared to the current state-of-the-art.


\vspace{0.3em}
\noindent \textbf{Improvement of Continuous Router.}
In the non-storage case, applying the continuous router results in an average fidelity improvement of up to 8.90x, as shown in the \emph{Our Fidelity (non-storage)} column, along with a significant reduction in execution time, as indicated in the \emph{Our $T_{exe}(\mu s)$ (non-storage)} column. This improvement stems from the continuous router, which effectively reduces both movement time and the number of transfers. The optimization effects are particularly pronounced in large-scale programs or benchmarks such as QAOA-regular3, QAOA-regular4, QAOA-random, and QFT, which involve a substantial number of CZ stages and collective movements, resulting in longer execution times and contributing to increased decoherence errors.


\vspace{0.3em}
\noindent \textbf{Improvement of Storage Zone.}
The integration of a storage zone significantly enhances fidelity, as reflected in the \emph{with-storage} column of Table~\ref{tab:main_table}, yielding an average improvement of 313.86x compared to the non-storage case. This benefit arises because the storage zone preserves non-interacting qubits with negligible decoherence, virtually eliminating excitation errors during Rydberg excitation. The impact becomes more pronounced as program size and circuit complexity increase, as demonstrated by the substantial improvements in large-scale benchmarks such as QAOA-random, QFT, BV, and QSim-rand, with enhancements up to $8.15 \times 10^4$x. These scenarios involve more Rydberg excitations and expose a greater number of non-interacting qubits to Rydberg excitations, leading to considerable excitation errors.

While introducing a storage zone does introduce additional overhead due to inter-zone movements, we effectively mitigate this overhead and still achieve up to a 2.32x reduction in execution time compared to Enola. This is made possible by our continuous router and stage scheduler, which minimize the cost of inter-zone movements.

\vspace{3pt}
\noindent \textbf{Reduction in Compilation Time.}
As shown in the \emph{$T_{comp}$ Improv.} column of Table~\ref{tab:main_table}, our framework delivers remarkable compilation time improvements of up to 213.5x compared to Enola. This improvement grows as program size increases, highlighting its effectiveness in handling larger-scale computations. While NAQC compilation optimization for NISQ applications is NP-hard~\cite{wang2024atomique, tan2022qubit, tan2024compiling}, we address this challenge through a near-linear heuristic algorithm that efficiently manages its complexities. In contrast, Enola relies on Maximum Independent Set solvers with higher time complexity, leading to significantly longer compilation times.

\subsection{Ablation Study}\label{subsec:Ablation}

In this section, we analyze the impact of individual components in our solution on each fidelity factor as the circuit scales increase across various benchmark circuits.
As illustrated in Fig. \ref{fig:fidelity ablation}, the green, blue, purple, yellow area represents two-qubit infidelity, excitation error, transfer infidelity, decoherence error, respectively. Our framework does not introduce additional two-qubit gates, so we target the reduction of the last three fidelity components: excitation error, qubit transfer infidelity, and decoherence error. We also evaluate the acceleration achieved through multi-AOD configurations and assess their impact on fidelity across different benchmarks.

\vspace{0.3em}
\noindent \textbf{Excitation Error Reduction.}
As shown in the blue sections of Fig.\ref{fig:fidelity ablation}, our framework eliminates excitation errors compared to Enola, thanks to the integration of a storage zone. This improvement is especially significant in the QSIM-rand and BV benchmarks, as illustrated in Fig.\ref{fig:fidelity ablation}(b) and \ref{fig:fidelity ablation}(e). In these benchmarks, the quantum circuits contain numerous CZ blocks, leading to a large number of Rydberg excitations. Additionally, each CZ block includes relatively few CZ gates, leaving many non-interacting qubits exposed to excitation errors. As a result, our framework’s ability to optimize excitation errors is particularly impactful in these cases, as clearly reflected in the with-storage graphs in Fig.~\ref{fig:fidelity ablation}(b) and \ref{fig:fidelity ablation}(e).


\vspace{0.3em}
\noindent \textbf{Decoherence Error Reduction.} 
As exhibited in the yellow part of Fig. \ref{fig:fidelity ablation}, significant reductions in decoherence errors are also observed, particularly in the QAOA-regular3, QFT, and VQE benchmarks. This improvement is primarily due to our continuous router, which minimizes redundant movement operations, thereby reducing decoherence time and, consequently, lowering decoherence errors. In the QAOA-regular3, QFT, and VQE benchmarks, the density and frequency of movement operations are relatively higher within a stage, which provides greater optimization potential for reducing decoherence errors.

\vspace{0.3em}
\noindent \textbf{Transfer Fidelity Enhancement.} 
As shown in the purple part of Fig. 6, there is also a noticeable reduction in transfer error, although the improvement in transfer fidelity is comparatively less pronounced compared to other fidelity metrics. This is primarily because the fidelity of individual transfer operations is already very high, at 99.9\%. The optimization of transfer fidelity is relatively evident in the QAOA, QFT, and VQE benchmarks. These benchmarks involve denser movement operations within each Rydberg stage, resulting in more frequent transfer operations. This configuration creates a larger optimization space for enhancing transfer fidelity.

\begin{figure}[h!]
    \hspace{-30pt}
    \includegraphics[width=0.8\linewidth]{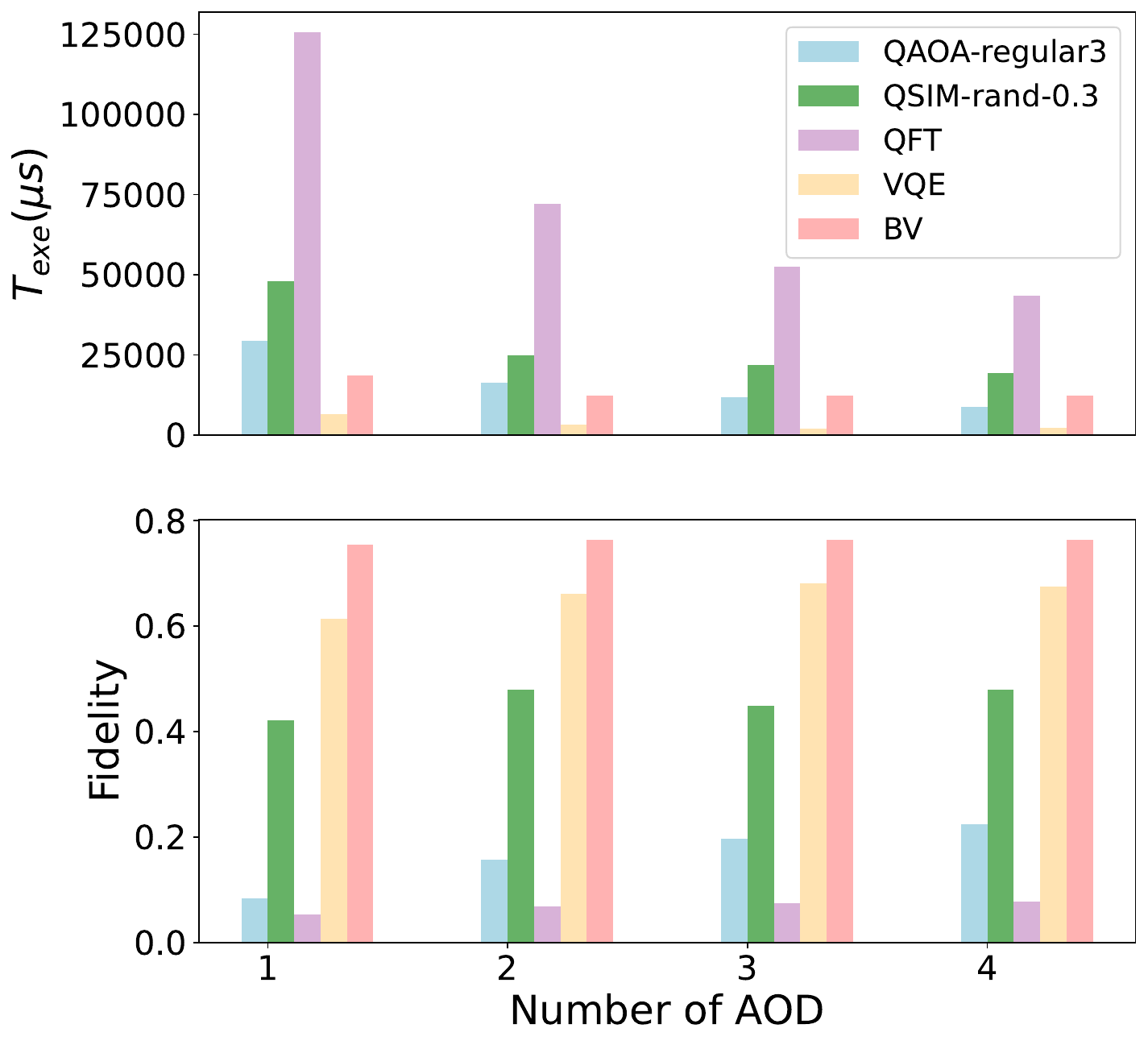}
    \caption{Effects of multiple AODs on the 100-qubit QAOA-regular3, 20-qubit QSIM-rand-0.3, 18-qubit QFT, 50-qubit VQE, and 70-qubit BV benchmark circuits.}

    \label{fig:multi_aod}
\end{figure}

\vspace{0.3em}
\noindent \textbf{Multiple AODs.}
Additionally, when we have less constrained hardware resources to utilize multiple AODs instead of a single one, we can leverage them to parallelize movement and transfer operations. As illustrated in Fig. \ref{fig:multi_aod}, even with a limited number of AODs, we achieve noticeable acceleration, demonstrating that substantial speed improvements can be realized without the need for excessive hardware resources. Notably, our fidelity optimization is more evident in benchmarks with higher decoherence errors, as shown in the fidelity section of Fig.~\ref{fig:multi_aod}, represented by the blue, green, and yellow bars corresponding to the QAOA-regular3, QSIM-rand-0.3, and VQE benchmarks, respectively. This improvement is due to the increased parallelism afforded by multi-AOD configurations, which effectively reduces decoherence time and, consequently, minimizes decoherence errors.

\section{Related Work}
\noindent \textbf{Quantum compiler in general.} Numerous compilation frameworks have been developed for specific quantum platforms, including superconducting compilers based on inserting SWAP gates~\cite{Qiskit, wille2023mqt, cowtan2019qubit, li2019tackling, zulehner2018efficient, tan2020optimal, omolecirq, Sivarajah2020tketAR}, photonic compilers based on fusion~\cite{zhang2023oneq, zhang2024oneperc, bartolucci2023fusion, bombin2021interleaving}, trapped ion compilers based on ion shuttling~\cite{pino2021demonstration, maslov2017basic, saki2022muzzle, groenland2020signal}, and neutral atom compilers leveraging atom movement~\cite{wang2024atomique, wang2023q, tan2022qubit, tan2024compiling, bochen2024compilation}, among others. Recent advances have also led to compilers for multi-chip architectures using inter-chip links~\cite{bravyi2022future, wang2018multidimensional} or distributed quantum computing~\cite{zhang2024mech, wu2022autocomm}. However, these compilers fail to exploit the specific capabilities of NA hardware, resulting in suboptimal performance when directly applied~\cite{tan2022qubit, tan2024compiling}. 

\vspace{3pt}
\noindent\textbf{Neutral atom compiler.} The development of compilers for NAQC has evolved in response to its continuously advancing hardware features. Early NA compilers focused on leveraging long-range CZ gates between atoms in static SLM traps within a limited radius~\cite{baker2021exploiting}, which extended neighborhood connectivity but suffered from low fidelity. Later approaches incorporated native multi-qubit gates like CCZ and proposed alternative qubit layouts~\cite{patel2022geyser} (Geyser), but were still constrained by limited long-range interactions. These early architectures, referred to as \emph{Fixed Atom Arrays (FAA)}~\cite{li2023timing, brandhofer2021optimal}, have gradually given way to \emph{reconfigurable atom arrays (RAA)}, or \emph{dynamically programmable qubit arrays (DPQA)}~\cite{tan2022qubit, nottingham2023decomposing}, with the advent of atom movement techniques~\cite{bluvstein2022quantum}. These newer compilers combine static SLM traps with mobile AOD traps, moving targeted qubits closer to perform parallel CZ gates. For interactions within SLM or AOD atoms, these compilers either switch trap types between SLM and AOD~\cite{tan2022qubit, tan2024compiling, bochen2024compilation} (OLSQ-DQPA, Enola), use SWAP gates to mobilize SLM qubits via AOD~\cite{wang2024atomique} (Atomique), or introduce ancilla qubits combined with CNOT gates for routing~\cite{wang2023q} (Q-Pilot). However, their performance is constrained by the significant overhead of atom movements and inserted two-qubit gates, or scalability issues.


\section{Conclusion}
In this paper, we present \textbf{\frameworkname}, a novel compilation framework for neutral atom quantum computers (NAQC) that fully leverages qubit movement capabilities while seamlessly integrating the newly developed Zoned Architecture (ZA). Our approach is the first to incorporate a storage zone into NAQC, effectively eliminating excitation errors with minimal overhead. We designed three key components that capitalize on the interplay between different aspects of the problem, efficiently navigating the vast design space to deliver solutions with high fidelity and scalability. Our evaluation demonstrates substantial improvements in output fidelity, alongside significant reductions in both execution time and compilation time. This work not only opens new avenues for optimizing NAQC compilation in NISQ applications through the use of ZA but also lays the groundwork for future compiler optimization in fault-tolerant quantum computing.



\section*{Acknowledgment}
We thank the anonymous reviewers for their constructive feedback and AWS Cloud Credit for Research. This work is supported in part by Robert N.Noyce Trust, NSF 2048144, NSF 2422169, NSF 2427109.
This material is based upon work supported by the U.S. Department of Energy, Office of Science, National Quantum Information Science Research Centers, Quantum Science Center. 
This research used resources of the Oak Ridge Leadership Computing Facility, which is a DOE Office of Science User Facility supported under Contract DE-AC05-00OR22725. The Pacific Northwest National Laboratory is operated by Battelle for the U.S. Department of Energy under Contract DE-AC05-76RL01830. 

\bibliographystyle{plain}
\bibliography{references}

\end{document}